\documentclass{aa}
\usepackage{graphicx}
\usepackage{graphics}
\usepackage{natbib}
\usepackage{subfigure}
\usepackage[usenames]{color}
\newcommand{\SiIII}{\ion{Si}{iii}}
\newcommand{\CI}{\ion{C}{i}}

\newcommand{\NeVI}{Ne\ts$\scriptstyle{\rm VI}$}
\newcommand{\PV}{P\ts$\scriptstyle{\rm V}$}
\newcommand{\CaX}{Ca\ts$\scriptstyle{\rm X}$}
\newcommand{\FeXII}{Fe\ts$\scriptstyle{\rm XII}$}
\newcommand{\FeXIX}{Fe\ts$\scriptstyle{\rm XIX}$}

\newcommand{\FeXXIV}{Fe\ts$\scriptstyle{\rm XXIV}$}

\newcommand{\OV}{O\ts$\scriptstyle{\rm V}$}

\newcommand{\Halpha}{H$\alpha$}
\newcommand{\kms}{km~s$^{-1}$}

\em
\usepackage{txfonts}

\begin{document}

   \title{Oscillations in the wake of a flare blast wave}

   \author{D. Tothova\inst{1}
\and D.E. Innes\inst{1} \and G. Stenborg\inst{2}}

\institute{ Max-Planck Institut f\"{u}r Sonnensystemforschung, 37191 Katlenburg-Lindau, Germany
\and Interferometrics, Inc. 13454 Sunrise Valley Drive, Herndon, VA 20171 , USA}

\institute{ Max-Planck Institut f\"{u}r Sonnensystemforschung, 37191 Katlenburg-Lindau, Germany}

   \offprints{D. Tothova \email{tothova@mps.mpg.de}}

  \date{Received ----; accepted ----}

\definecolor{orange}{cmyk}{0,0.5,1,0}

\abstract
{Oscillations of coronal loops in the Sun have been reported in both imaging and spectral observations at the onset of flares.
Images reveal transverse oscillations, whereas spectra detect line-of-sight velocity or Doppler-shift oscillations. The Doppler-shift oscillations are commonly interpreted as longitudinal modes.}
{Our aim is to investigate the relationship between loop dynamics and flows seen in TRACE 195\AA\
images and Doppler shifts observed by SUMER in \SiIII\ 1113.2\AA\ and \FeXIX\ 1118.1\AA\ at
the time of a {C.8}-class limb flare and an associated CME.}
 {We carefully co-aligned the sequence of TRACE 195\AA\ images to structures seen in the SUMER \SiIII, \CaX\ ,and \FeXIX\ emission lines. Additionally, \Halpha\ observations of a lifting prominence associated with the flare and the coronal mass ejection (CME) are available in three bands around 6563.3\AA. They give constraints on the timing and geometry. }
{Large-scale Doppler-shift oscillations in \FeXIX\ and transverse oscillations in intensity images were observed over a large region of the corona after the passage of a {wide bright extreme-ultraviolet (EUV) disturbance}, which suggests ionization, heating, and acceleration of hot plasma in the wake of a blast wave.}
{}


\titlerunning{}
 \authorrunning{D. Tothova}

 \maketitle

\section{Introduction}

Oscillations in flare loops on the Sun are caused by a rapid injection of energy to the plasma in the
loop. Kink oscillations, manifested as periodic transverse displacements of coronal loops, are
believed to be triggered by a flare blast wave that hits the loop edge-on \citep{Aschwanden:1999rw,
Aschwanden:2002ek, Nakariakov:1999it, Schrijver:2002qq}. In their study, \citet{Hudson:2004gf} found that 12 out of 28 cases of TRACE oscillations concide with type II bursts, which supports their strong connection with large-scale flare shocks.

We present a detailed study of SUMER spectroscopic images, revealing high-velocity blue and red Doppler shifts in lines of \SiIII\ and \FeXIX\ before, during, and after {the passage of a large concentric disturbance in  195\AA\ images.} 

{This disturbance is followed by a filament eruption and large-scale oscillations of the ambient plasma.} The high-velocity shifts, which are significant far beyond the edge of the rising prominence, are initially observed in the chromospheric \SiIII\ line. When a rapid increase in the intensity of \FeXIX\ and a simultaneous rapid drop in the \SiIII\ intensity occur, the oscillation is seen in the flare line \FeXIX. We propose that the
observations show ionization, heating, and acceleration of
plasma in the wake of a flare blast wave.

{By the term `blast wave' we mean `a large-amplitude disturbance which propagates as a nonlinear simple wave'  \citep{Mann:1995aa,Nindos:2008aa}, i.e. a  fast magnetosonic wave with a large amplitude. We use the term `wide front' to describe the propagating bright wide disturbance seen in TRACE 195\AA\ (Fig.~\ref{movie} and \ref{TRACE_h_slits}) followed by dimming. This should not be confused with a shock front as observed in density-sensitive white light images \citep{Vourlidas:2003aa}. The emission in extreme-ultraviolet (EUV) is strongly temperature-sensitive, so that if there were density changes, these would not be detected. } 

\section{Observations}

\begin{figure}
\centering
\includegraphics[width=3.15in]{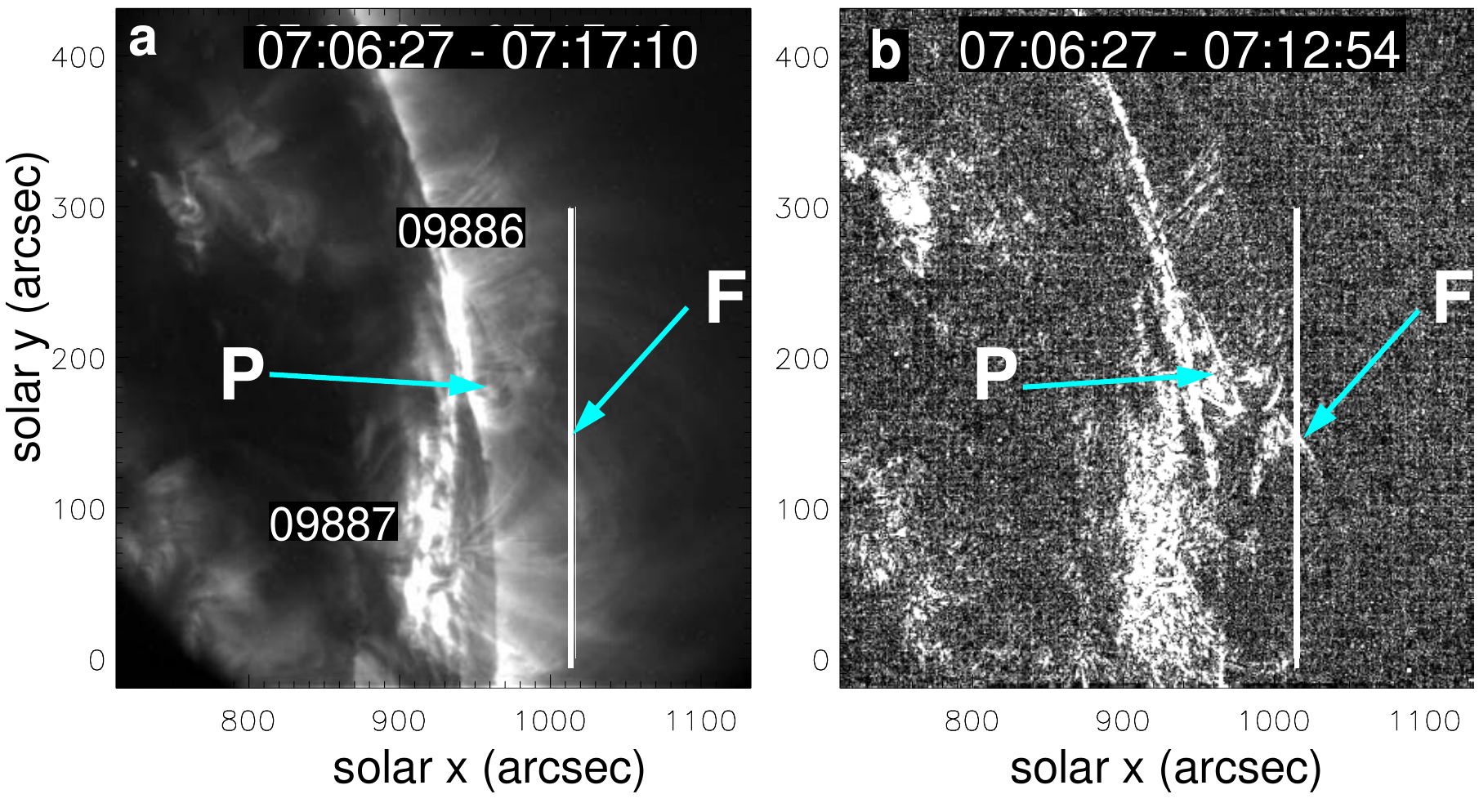}
\caption{a) Average TRACE 195\AA\ image of AR09886 and surrounding region: (a) average intensity between 07:06 and 07:17~UT showing the prominence (P) that later erupted; (b) power map for frequencies of less than 5mHz (periods greater than 3.3 min). Flows (F) are detected in the corona crossing the SUMER slit, which is indicated with a white vertical line.}
\label{tr_over}
\end{figure}

{The C.8 class flare occurred around 07:24~UT on 9 April 2002 in AR09886 near the western limb of the Sun  (Fig.~\ref{tr_over}a)}. Faint coronal loops can be seen extending high into the corona. Some connect to AR09887 {on} this side of the limb, and others seem to be directed toward active regions, AR09885 and AR09891, already behind the limb. We show in Fig.~\ref{tr_over}b the power map of the lowest frequency intensity variations in the pre-flare phase. This reveals activity in the prominence (P) and flows (F) in the corona.

\begin{figure}
\centering
\includegraphics[width=3.15in]{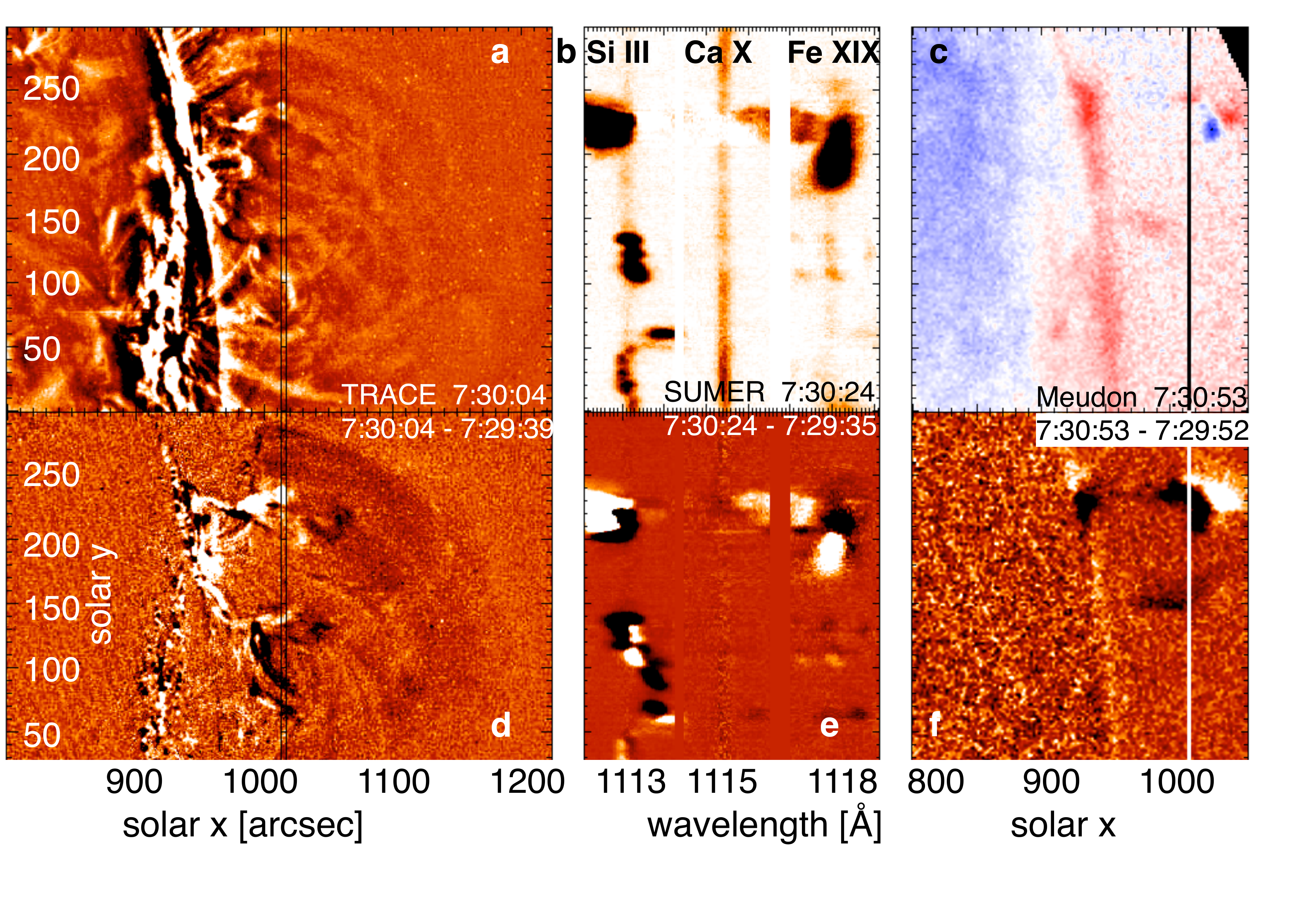}
\caption{Flare evolution: (a) TRACE contrast-enhanced intensity (b) SUMER spectral windows centered at 1113\AA, 557\AA , and 1118\AA\ shown in reverse color (c) Meudon heliograph \Halpha\ Dopplergrams (d) TRACE running difference (e) SUMER running difference (f) \Halpha\ running difference. Two vertical lines indicate the position of the SUMER slit in the TRACE and \Halpha\ images. This is one frame of the online movie {\tt 02apr09\_movie.pdf}.}
\label{movie}
\end{figure}

An overview of the evolution of the loop system seen in TRACE, SUMER, and \Halpha\ is shown in the online movie associated with Fig.~\ref{movie}, starting at 07:06~UT, the beginning of the TRACE observations.
The movie shows simultaneous TRACE and SUMER intensity images, and \Halpha\ Dopplergrams in the top row. Below are the corresponding running difference intensity images.
Each frame is scaled individually, thus hiding the time evolution of the intensity, which is represented in TRACE and SUMER time-distance images (Figs.~\ref{TRACE_h_slits} and \ref{SUMER_slits}, respectively).

TRACE \citep{Hetal98} observed in the 195\AA\
band with a cadence of roughly 13~s.
The emission is caused by \FeXII\ (1.6~MK) in the quiet Sun, and is dominated by emission
from hotter \FeXXIV\ 193\AA\ (20~MK) under the conditions
found in flaring active region coronae.
Coronal dimming at 195\AA\ {may therefore be interpreted as either cooling, heating,} absorption by hydrogen and helium along the line-of-sight, or a decrease in plasma density.
In the filter band the transition region lines \OV\ 192.8 and 192.9\AA\ \citep{Young07b} may contribute to the observed prominence emission.
{The TRACE data were averaged} within each 50~s exposure period of SUMER, so that they overlap as much as possible in time.
For the intensity movie (Fig.~\ref{movie}a) the averaged images were additionally contrast-enhanced by subtracting a blurred image obtained by using repeated convolutions of a bi-dimensional mask \citep{Stenborg:2003fk} on the image. The running difference movie uses the averaged data.

{ \Halpha~images} were obtained with the Meudon heliograph at a 1~min cadence in the core (6563.3\AA) and in the wings ($\pm 0.5$\AA) of the \Halpha\ line. The Doppler shift, Fig.~\ref{movie}c, is computed as the first weighted moment of the {emissivity.} The maximum measurable shift of 0.5\AA, when all the emissivity comes from one wing of the line, yields a maximum measurable velocity of about 23~\kms. This is much lower than the velocities measured with the SUMER spectrometer. Any \Halpha\ emission coincident with the high-velocity \SiIII\ plasma seen during the eruption was outside the filter window.

From around 07:20~UT, a prominence is seen to rise in the \Halpha\ intensity running difference images (Fig.~\ref{movie}f), showing faint Doppler-shift signatures. The apparent plane-of-sky velocity of the rising prominence
was in the range $120-160$~\kms, which is significantly lower than the fast moving features seen in SUMER data mentioned below. 

The SUMER spectrometer \citep{Wilhelm:1995uq} observed in sit-and-stare mode
about 50\arcsec\ off
the west limb, which allows the observation of loops formed within and among active regions NOAA AR09886, 09887, 09885, and 09891 with the 300\arcsec x4\arcsec\ slit and a 50~s
cadence simultaneously in the three lines: \SiIII\ 1113.24\AA\ (0.06~MK), \CaX\ 557.76\AA\
(second order, 0.7~MK) and \FeXIX\ 1118.1\AA\ (6~MK). The line-formation temperatures are given in parenthesis.
The size of the spatial and spectral pixels was about 1\arcsec\
and 44~m\AA\ respectively. The width of the spectral window around each line was 50~px, corresponding
to about 2.2\AA, or in terms of the line-of-sight component of the velocity, to approximately
590~\kms. 

{In} Fig.~\ref{movie}b,e and the movie the three spectral windows \SiIII\ , \CaX\ , and \FeXIX are plotted side by side with data gaps of 50~\kms\ and 140~\kms\ between the windows.

The \FeXIX\ 1118.07\AA\ line blends with both the transition region, \PV\ 1117.98\AA\ and \NeVI\ 558.62\AA, and the cooler \CI\ lines at 1117.20, 1117.58, 1117.88, 1118.18, and 1118.49\AA\ \citep{Cetal01}. We used the \SiIII\ to identify potential transition region blending. For example, in Fig.~\ref{movie}b, the broad, strong emission around $y$=230\arcsec\ , which coincides with strong \SiIII\ , is \NeVI, but the emission below, between $150-200$\arcsec, is \FeXIX. Here high \NeVI\ Doppler-shifted emission is visible beyond the 140~\kms\ data gap in the \CaX\ window.
The \CI\ lines can be identified because another \CI\ multiplet (1114.64, 1114.86 and 1115.21\AA) appears at the same time in the \CaX\ window.
Our analysis only considers \FeXIX\ at times and positions when there is no strong \SiIII\ or \CI.

{Key} features used in the co-alignment were the position of the early pre-flare flows visible in TRACE and SUMER, the blue-shifted \NeVI, \SiIII\ , and \Halpha, visible as erupting prominence material in TRACE as well as  bright TRACE and red-shifted SUMER \SiIII\ and \Halpha\ along the northern leg of the erupting prominence at the end of the movie.
SUMER was found to be centered at (1015\arcsec, 150\arcsec), with an accuracy of 5\arcsec.

\subsection{Ejecta, fronts, and oscillations}

\begin{figure}
\centering
\includegraphics[width=3.15in]{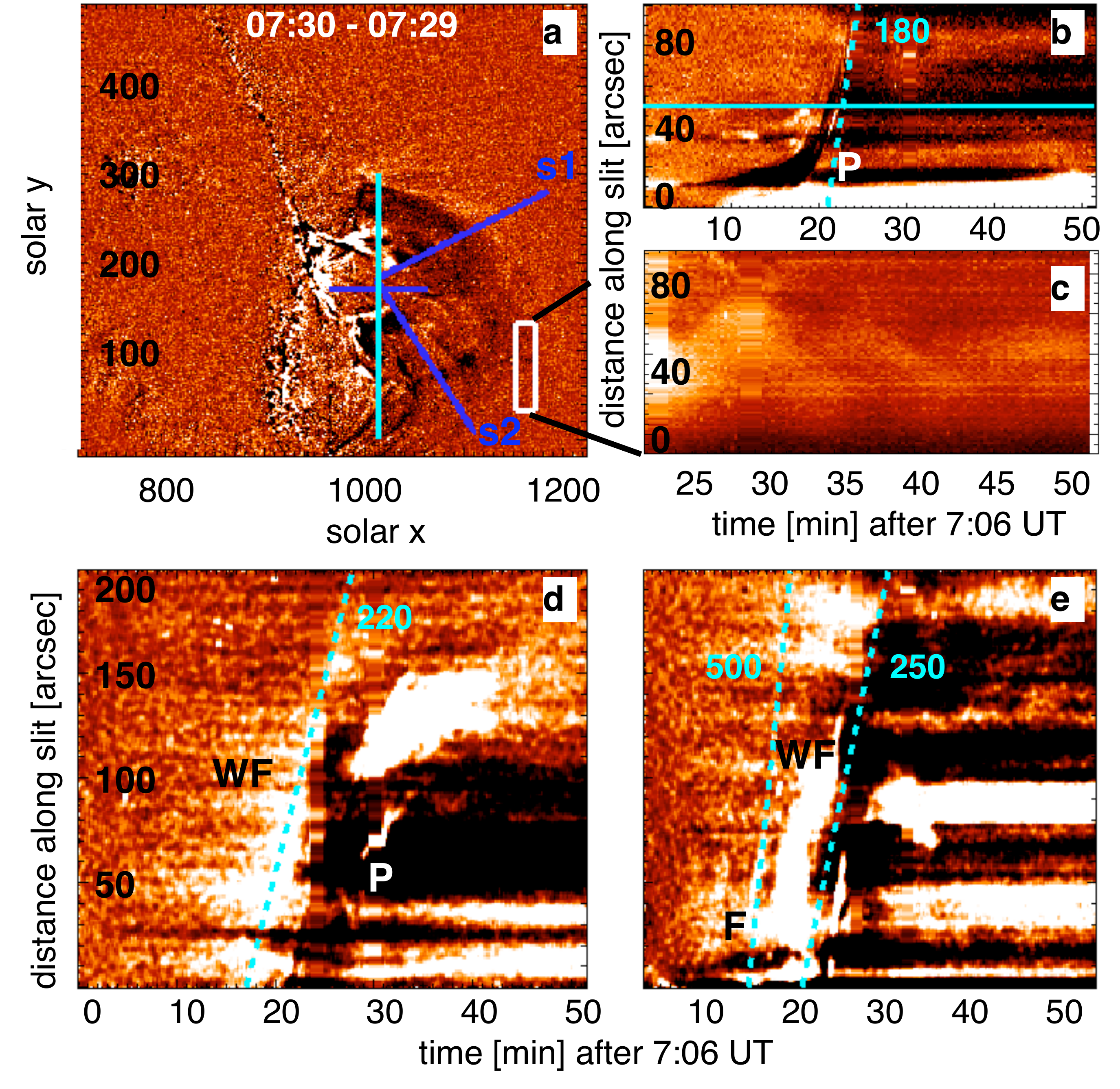}

\caption{{a) TRACE 195\AA\ running-difference context image with the synthetic slits (blue solid lines and the white rectangle). The position of the SUMER slit is represented with the vertical cyan line. b-e) TRACE 195\AA\ time-distance plots along the synthetic slits. The prominence (P), wide front (WF) and flows (F) are marked. Pre-event background intensities were subtracted from each time series. The apparent plane-of-sky velocities along the slits are given alongside the cyan dashed lines running along the erupting filament (b) and along the fronts and flows (d and e).
b) Intensity variation along the horizontal slit  constructed at $y$=170\arcsec. c)  Time evolution of TRACE intensity, averaged over the solar x width of the white rectangle in (a). d, e) Intensity variation along the slits labelled s1 and s2, respectively. }}
\label{TRACE_h_slits}
\end{figure}

The disruption of the corona can be seen in the TRACE {movie.} 
To obtain the {the lower limit of the} expansion speeds of some of the
observed fronts {and flows}, we constructed time-distance intensity images {along synthetic slits (Fig.~\ref{TRACE_h_slits}b, d, e}).
Bright fronts {and flows} are visible throughout the region. 
They are associated with the eruption that started along $y$=170\arcsec\ at 07:20~UT (Fig.~\ref{TRACE_h_slits}{b}) in which both prominence and corona take off simultaneously, similar to the mini-CME onsets reported by \citet{Innes:2010aa}.
The {prominence, visible as a dark front, reached} the SUMER slit at 07:2{9}~UT. 

{In the time-distance plots in Fig.~\ref{TRACE_h_slits}d, e, which were constructed along the slits s1 and s2 (represented in Fig.~\ref{TRACE_h_slits}a), the wide front (WF) precedes the prominence and the dimming. It reaches much greater heights than the expanding filament. The velocities given beside the  wide front are a guide to the plane-of-sky speed. The actual speed may be faster because weak emission enhancement is visible high in the corona at earlier times.  Flows (F), which are marked in Fig.~\ref{TRACE_h_slits}e, were triggered in loops (see the movie) before the prominence eruption. We suggest that they are triggered by a blast wave.}

Striking in the TRACE intensity frames of the online movie, but difficult to show with stills, is the large-scale back-reaction of the loop systems to the passing front. By making space-time images along different tracks in the corona, we were able to pick out the displacement oscillation shown in Fig.~\ref{TRACE_h_slits}c. It looks as though a broad, 20\arcsec\ wide structure was oscillating up and down in latitude with a period of about {8~min}. 

\begin{figure}
\centering
\includegraphics[width=3.15in]{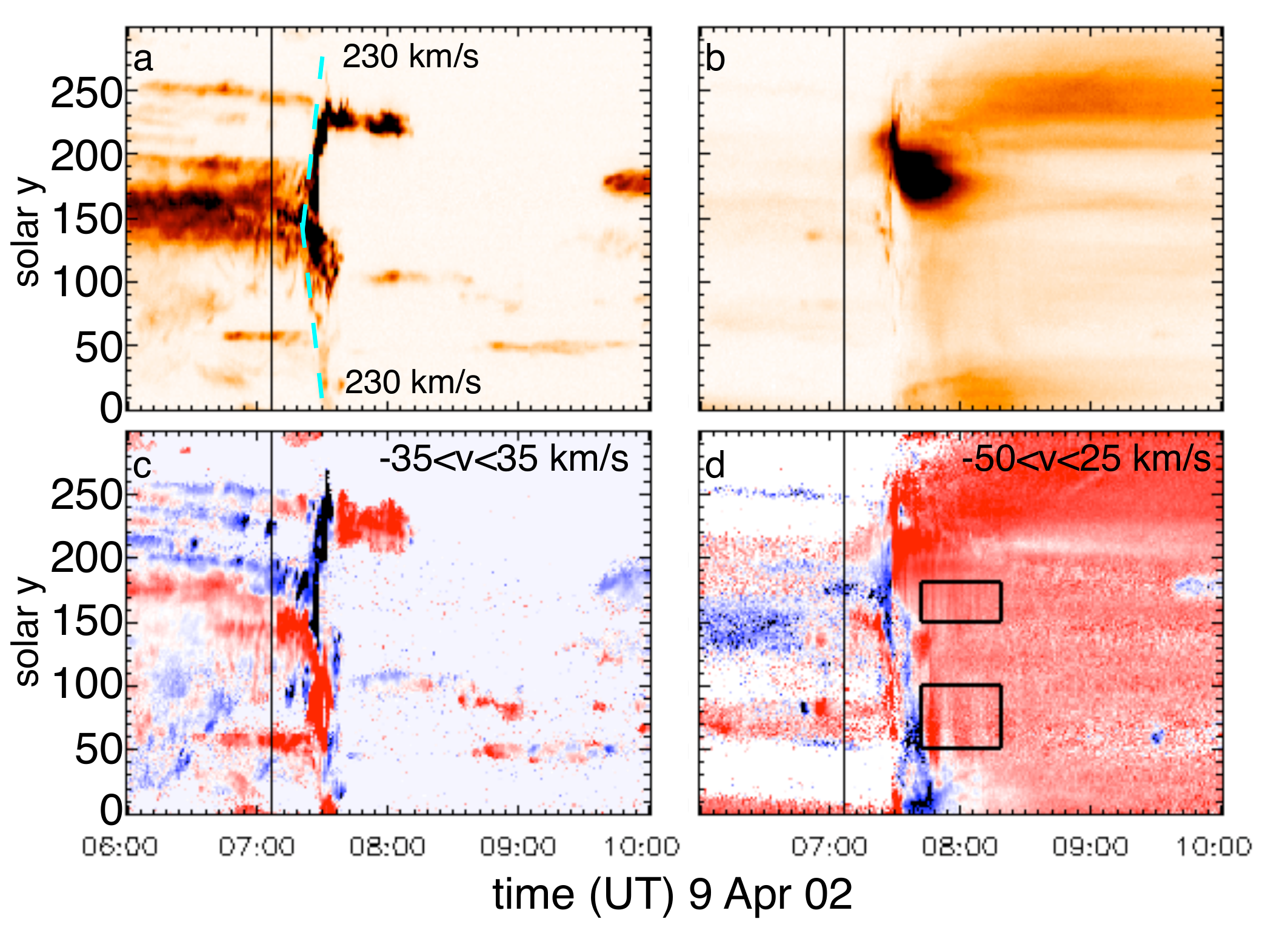}

\caption{SUMER line intensities and Doppler shifts: (a) \SiIII\ intensity (b) \FeXIX\ intensity (c) \SiIII\ Doppler shift (d) \FeXIX\ Doppler shift. Intensities are reverse color. The start time of the movie is marked with a black vertical line. The rectangles in (d) outline the oscillation regions shown in Figs.~\ref{wavelet_fig} and \ref{oscill_fig}.}
\label{SUMER_slits}
\end{figure}

The SUMER \SiIII\ and \FeXIX\ time series in Fig.~\ref{SUMER_slits} show the spatial development of the transition region and hot flare plasma along a single vertical slit in the corona. The spectral details are best seen in the online movie. The \SiIII\ red shifts in the middle of the SUMER slit correspond to velocities faster than 300~\kms. The early flows, directed southward and away from the observer, coincide with the coronal flows (F) seen in {Fig. \ref{TRACE_h_slits} e) and in} the TRACE power map (Fig.~\ref{tr_over}b).
 At 07:20~UT, the time of the two-tier eruption (Fig.~\ref{TRACE_h_slits}{b}), SUMER detected the coronal part of the disturbance. The transition region plasma moved rapidly to the north and south with an apparent plane-of-sky velocity of 200~\kms.

 The large \SiIII\ red shifts seen later along the southern part of the SUMER slit show up as flows in the TRACE running difference images of the online movie. The images suggest that the movement of \SiIII\ along the slit was not plasma motion, but marks a front triggering \SiIII\ flows. In the north, we can associate the 450~\kms\ \SiIII\ and \NeVI\ blue shifts seen in SUMER at 07:30~UT (Fig.~\ref{movie}) with erupting prominence plasma. Immediately after the erupting prominence, the \FeXIX\ intensity increased. It was particulary bright above the active region, but significant emission also appeared to the south along the slit. Initially, the \FeXIX\ line was predominantly blue-shifted with a non-Gaussian wing extending to 150~\kms.

\begin{figure}
\centering
\includegraphics[width=3.15in]{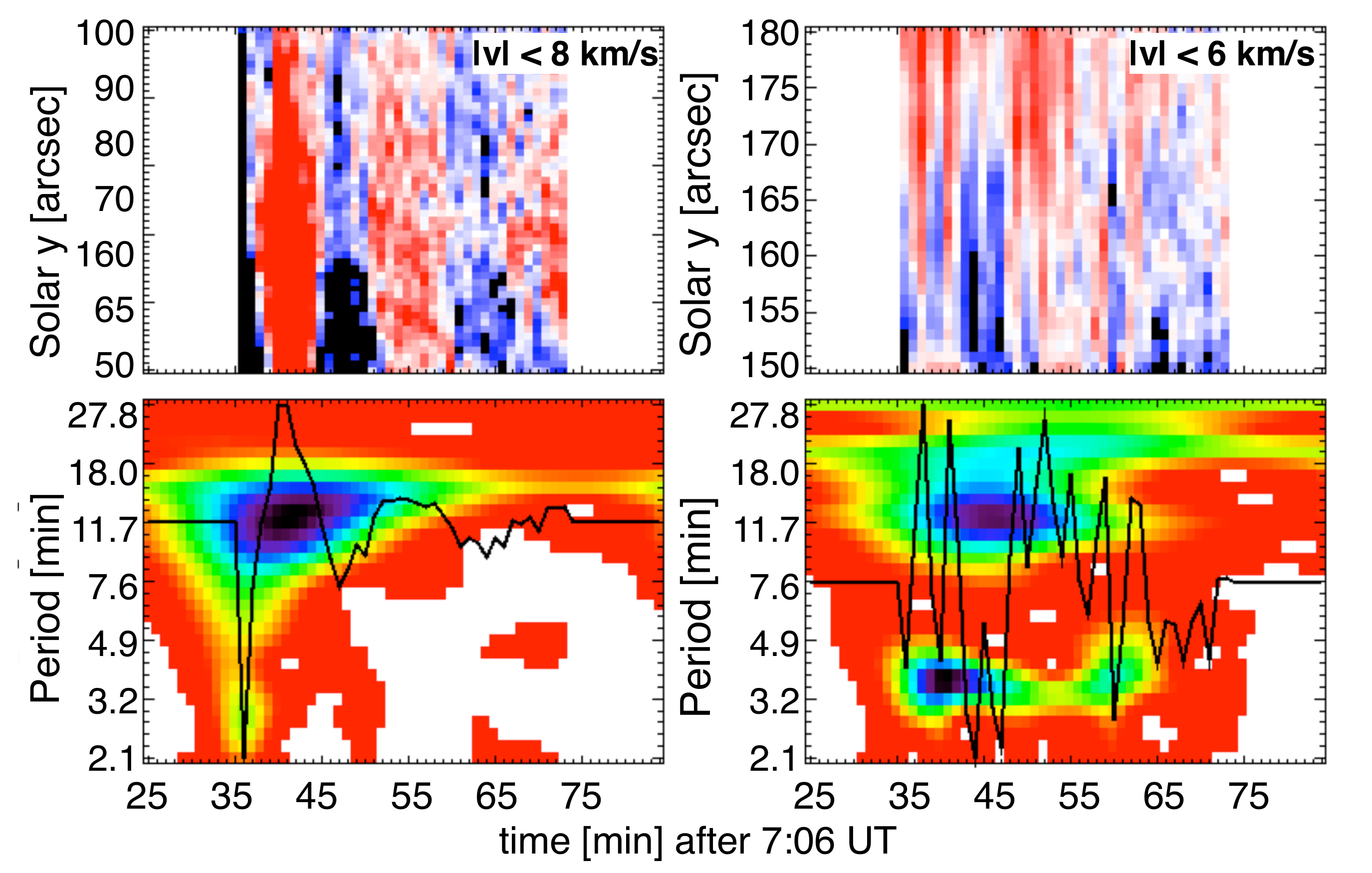}
\caption{Detail of the Doppler-shift oscillation in \FeXIX\ for (a) the lower rectangle in Fig.~\ref{SUMER_slits}d and (b) the upper rectangle in Fig.~\ref{SUMER_slits}d. The frames below show the wavelet power of the Doppler shifts in the frame above with the average Doppler shift overplotted.}
\label{wavelet_fig}
\end{figure}

Weak oscillations following the front are visible in the \FeXIX\ space-time Doppler shift images (Fig.~\ref{SUMER_slits} ).
Details of the two {regions}, which are outlined by black rectangles in Fig.~\ref{SUMER_slits}d, are shown in Figs.~\ref{wavelet_fig}a and b.  The time span of the details is 33 minutes and is represented by the length of the horizontal  side of the rectangle. In Figs. \ref{wavelet_fig}c and d we show the wavelet power of the line center Doppler shift averaged over the 50\arcsec\ and 30\arcsec\ regions of the slit shown in the figure.
The one to the south at $50\arcsec < y < 100\arcsec$ has a single period of about 14~min. The other at $150\arcsec < y < 220\arcsec$ is more complex. It is composed of two periods, one around 4~min and the other also around 14~min. No oscillation is seen in the region between.

\begin{figure}
\centering
\includegraphics[width=3.15in]{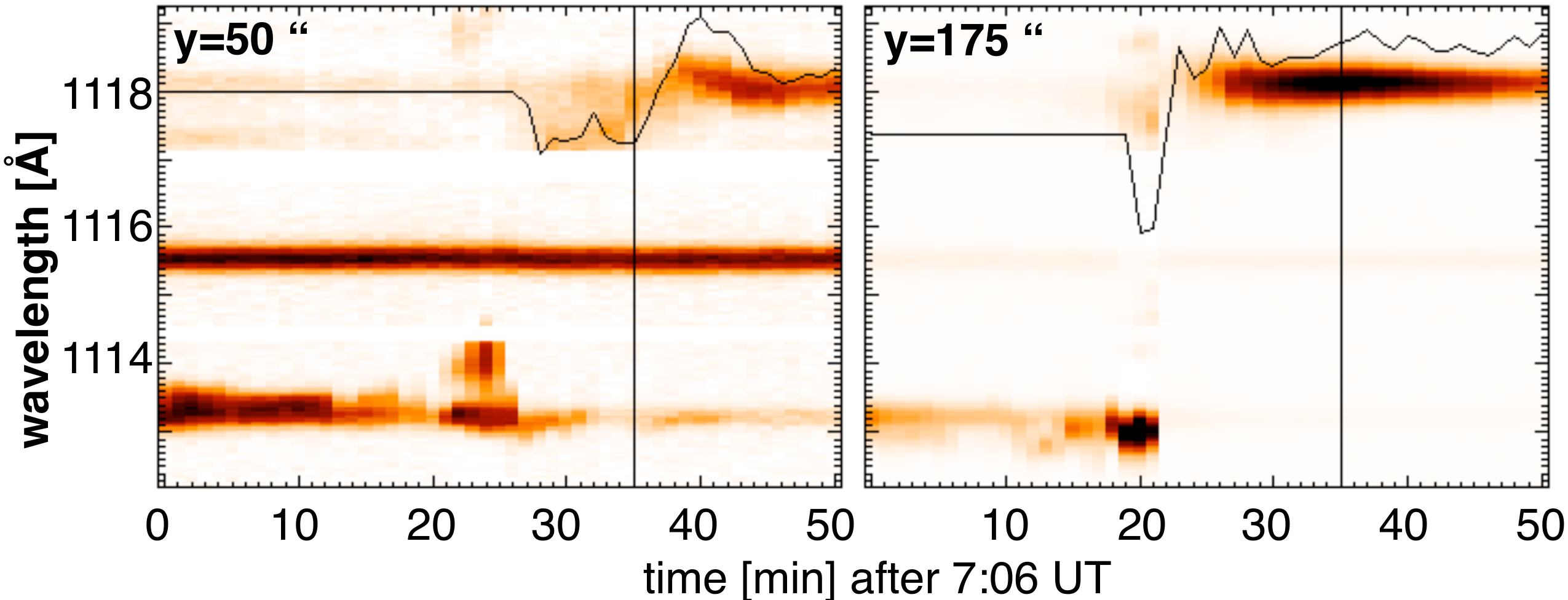}
\caption{Spectral evolution at positions of the Doppler-shift oscillations. The spectra were averaged over 10\arcsec\ and displayed with the \FeXIX\ window at the top, \CaX\ in the middle and \SiIII\ at the bottom. A black vertical line was drawn at the start of the measured \FeXIX\ oscillations.}
\label{oscill_fig}
\end{figure}

In Fig.~\ref{oscill_fig} we plot the time evolution of the spectra averaged over 10\arcsec\ and centered at the strongest points of the oscillations. Vertical black lines denote the start of the oscillations time span, 7:42 - 8:22. The transition region \SiIII\ emission disappeared before the \FeXIX\ oscillations, and we conclude that the oscillation is definitely caused by the shifts in \FeXIX\ line and not by contamination by its blends.
The spectral evolution shown in Fig.~\ref{oscill_fig}a is typical for the positions $50\arcsec < y < 100\arcsec$ in which we see several such \SiIII\ jets followed by an \FeXIX\ oscillation starting with a blue shift. This seems to support the impression given by the images that the {flows and oscillations} were triggered by a front moving through the corona, because a large-scale front would also heat {and ionize} plasma in its wake. It is unclear whether the \FeXIX\ blue shift was caused by a back reaction to the red-shifted \SiIII\ jets or if it was different plasma accelerated behind a part of the front moving toward the Earth.

\section{Discussion and conclusions}


{We presented EUV spectroscopic and imaging observations of a flare accompanied by a large-scale wave, an oscillation in its wake, and a prominence eruption. Here we summarize the arguments for the interpretation of the data in terms of ionization, heating, and acceleration in the wake of a blast wave.}

{A wide bright front was seen ahead of the erupting filament in the TRACE 195\AA\ images. Its width comes from the delay caused by the finite duration of ionization. Owing to the optical thinness of the 195\AA\  line, the projection effects may also play a role. As it passed through the field-of-view of the} SUMER slit, the \SiIII\ line brightened and brief bursts of red-shifted emission were visible from positions moving south and north along the slit. The \SiIII\ intensity then decreased rapidly. At the same time, the intensity of the flare line \FeXIX\ increased and a Doppler-shift oscillation lasting at least 1 hour was launched along broad sections of the slit, with a varying amplitude and period.

  { The disappearance of \SiIII\ was either caused by {ionization and heating,} or by ejecta moving out of the slit field-of-view. Initially, TRACE shows many bright ejecta moving south intermingled with a background dimming.} Afterward we see {a long-lasting} increased \FeXIX\ emission, which implies heating, and oscillating loops across the region, which implies plasma acceleration on a large scale. We therefore suggest that the \SiIII\ flows are triggered by a front moving through the corona.

We detected at least three different oscillations with SUMER and TRACE at different positions. {From their timing it seems very probable that they have the same trigger.} SUMER shows large regions with 14~min oscillations at $50\arcsec < y < 100\arcsec$ and $150\arcsec < y < 180\arcsec$. A shorter period (4~min) oscillation was superposed on the 14~min one in the northern section. A TRACE oscillation with a period of {8}~min was picked up about 100~Mm higher in the corona. It may have been present closer in, but there is lots of confusion by overlapping structures below $x$=1100\arcsec\ , so it is very difficult to see.
All oscillations showed a high initial pulse, which supports the idea of an impulsive trigger {by a large-amplitude wave}.

Doppler oscillations in \FeXIX\ of this width and with periods about 14~min are commonly observed at the onset of flares \citep{Wetal03}. They are generally interpreted as caused by slow-mode standing waves
 in hot coronal loops excited at their footpoints, because their periods match those expected for the fundamental of the slow mode. This is true for the loops here as well. If we assume a semi-circular loop of height 50~Mm (the height of the slit from the limb) and a sound speed 380~\kms  \citep{Wetal03}, the fundamental mode would have a period of 14~min.
 The difference between the events in \citet{Wetal03} and the event described here is that here there were (i) many, not just one loop, along the line-of-sight, (ii) the trigger seems to be the same as the one that caused the TRACE oscillation, and (iii) there was a short (4~min) period Doppler oscillation that cannot be attributed to a slow-mode standing wave.The large scale of the oscillations also suggests that the excitation site was in the corona rather than at a loop footpoint.

 In 1-D loops, a coronal trigger would excite the first harmonic  of the slow mode,  and its expected period would be half the fundamental (i.e. $\sim$ 7~min), which is too short to explain the observed oscillation. However, when a twisted loop erupts, \citet{Haynes:2008aa} demonstrated with simulations of a 3-D loop that the first harmonic of the slow mode rapidly relaxes to the fundamental slow-mode wave. Thus, the large-scale 14~min
 oscillations could be the fundamental slow mode triggered by a passing blast wave in the corona. The 4~min oscillation could then be explained as the Doppler signature of the transverse kink oscillation in loops at the height of the SUMER slit. The {8~min} TRACE oscillation was observed at almost twice the height in the corona, and can therefore be explained as the fundamental kink mode in a loop with twice the length.

The arguments mentioned above lead us to consider heating and acceleration in the wake of a large-scale blast wave flowing around magnetic obstacles - coronal loops - formed within and among the active regions covering a large region on the Sun as the cause for the loop oscillations.

\begin{acknowledgements}
SUMER is financially supported by DLR, CNES, NASA, and the ESA PRODEX program
(Swiss contribution). SoHO is a project of international co-operation between
ESA and NASA.
\end{acknowledgements}

\bibliographystyle{aa}


\end{document}